\documentclass[aps,reprint,pre,superscriptaddress,floatfix]{revtex4-2}

\usepackage{amsmath}
\usepackage{graphicx}
\usepackage[separate-uncertainty]{siunitx}
\usepackage[version=4]{mhchem}
\usepackage{tabularx}

\graphicspath{{figures/}}

\DeclareSIUnit{\Molar}{M}
\DeclareSIUnit{\mM}{\milli\Molar}

\begin{document}

\title{Multiplexed holographic molecular binding assays with internal calibration standards}

\author{Kaitlynn Snyder}

\author{Andrew D. Hollingsworth}
\affiliation{Department of Physics and Center for Soft Matter Research, New York University, New York, New York 10003, USA}

\author{Fook Chiong Cheong}
\affiliation{Department of Chemistry, New York University, New York, New York 10003, USA}

\author{Rushna Quddus}
\affiliation{Adolph Merkle Institute, University of Fribourg,
CH-1700 Fribourg, Switzerland}

\author{David G. Grier}
\affiliation{Department of Physics and Center for Soft Matter Research, New York University, New York, New York 10003, USA}

\date{\today}

\begin{abstract}
Holographic molecular binding assays detect
macromolecules binding to colloidal probe beads
by monitoring nanometer-scale changes in the beads' diameters with holographic microscopy.
Measured changes are interpreted with Maxwell Garnett
effective-medium theory to infer the surface coverage of analyte
molecules and therefore to measure
the analyte concentration in solution.
The precision and accuracy of those measurements can be
degraded by run-to-run instrumental variations, which introduce
systematic errors in the holographic characterization measurements.
We detect and mitigate these errors by introducing a class of
inert reference beads whose polymer brush coating resists
macromolecular binding.
The holographically measured diameter and refractive index of those beads serve
as internal standards for THC measurements.
To characterize the reference beads, we
introduce a general all-optical method to measure the
grafting density of the polymer brush that combines
holographic characterization of the bead diameter with
a refractometry measurement of the polymer's specific volume.
The latter technique shows the specific volume
of poly(ethylene oxide) to be
\qty{1.308(4)}{\cubic\nm\per\kilo\dalton}.
We use this suite of
techniques to demonstrate
a multiplexed immunoassay for immunoglobulin G (IgG) whose
success validates the effective-medium analysis of holographic
characterization measurements.
Internal negative controls provided by the reference beads
are validated by negative control measurements
on alcohol dehydrogenase (ADH), which has a similar
molecular weight to IgG but does not bind to the probe
beads' binding sites.
\end{abstract}

\maketitle

\section{Holographic molecular binding assays}

\begin{figure}
    \centering
    \includegraphics[width=0.8\columnwidth]{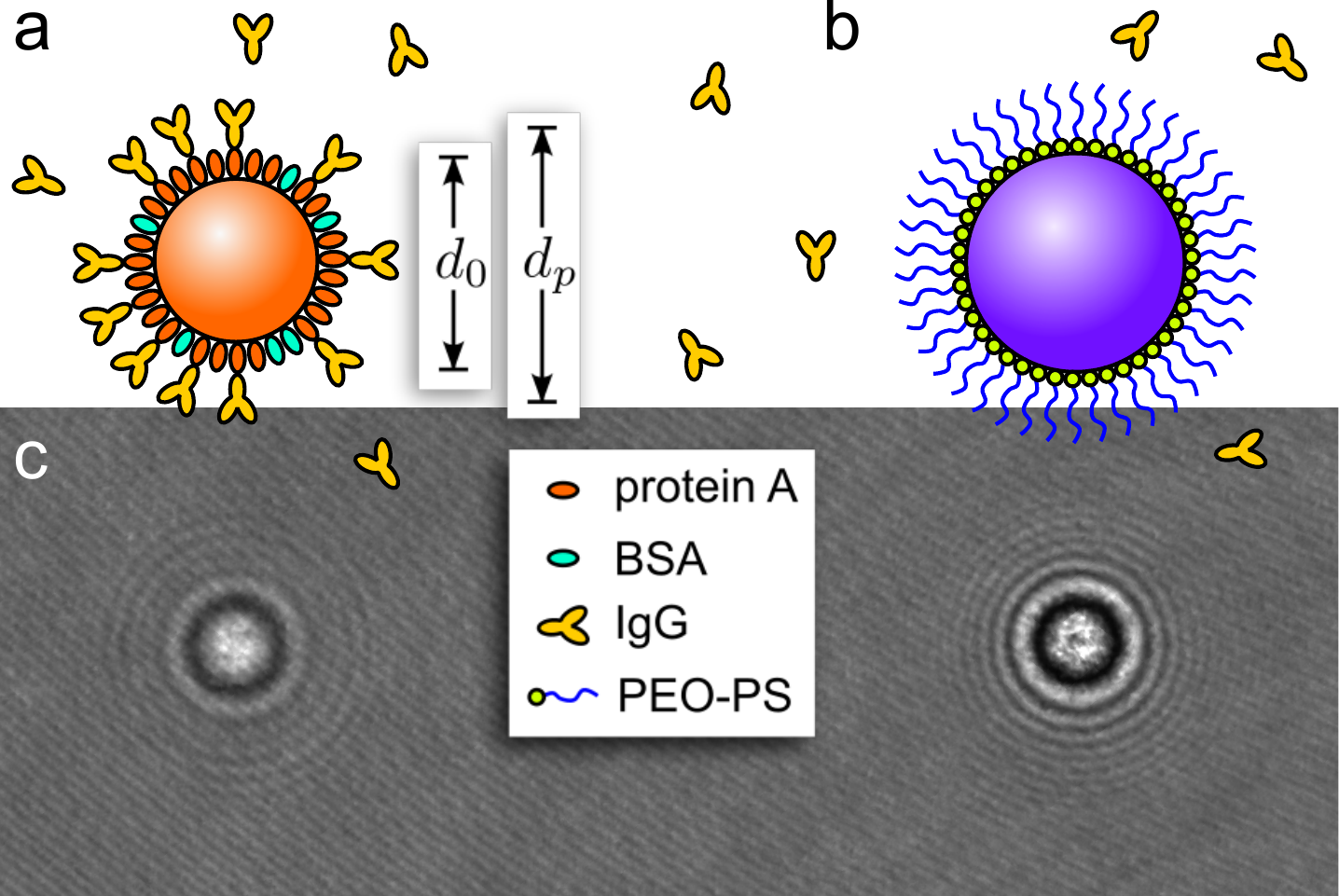}
    \caption{(a) A probe bead for a holographic immunoassay consists of a substrate bead functionalized with binding sites such as protein A.
    Bare spots on the surface are passivated with BSA.
    The hologram of such a bead encodes
    its starting diameter, $d_0$.
    Target analytes such as IgG
    bind to the functionalized bead,
    increasing its measured diameter to $d_p$.
    (b) The surface of a reference bead is covered with a dense PEO brush that inhibits macromolecular binding.
    Such inert beads serve as negative controls in
    binding assays.
    (c) In-line hologram of a silica-core probe bead
    (left) next to a PS-core reference bead (right).
    The beads are distinguishable by size and refractive index.}
    \label{fig:beads}
\end{figure}

Holographic molecular binding assays \cite{zagzag2020holographic,altman2020interpreting,snyder2020holographic}
use Total Holographic Characterization (THC)
\cite{lee2007characterizing}
to detect molecules binding to the surfaces of suitably functionalized
colloidal probe beads and thus to infer
the concentration of those
target molecules in solution.
THC is a label-free particle-characterization method
that records and analyzes holograms of individual
colloidal particles in their native environment.
When applied to micrometer-scale spheres, this
analysis yields each bead's diameter, $d_p$, with nanometer precision and therefore
can detect changes in diameter
due to analyte molecules accumulating on the surface \cite{lee2007characterizing,krishnatreya2014measuring,zagzag2020holographic,altman2020interpreting}.
THC simultaneously resolves each bead's refractive index,
$n_p$, with part-per-thousand precision
\cite{shpaisman2012holographic,krishnatreya2014measuring}.
The refractive index is useful for differentiating
different classes of probe beads by their composition and therefore is useful
for developing multiplexed bead-based binding assays.

A holographic binding assay, illustrated schematically
in Fig.~\ref{fig:beads}, compares
THC results for ensembles of probe beads before and after incubation with samples that
contain unknown concentrations of target molecules \cite{cheong2009flow,zagzag2020holographic,snyder2020holographic,altman2020interpreting}.
The diagram in Fig.~\ref{fig:beads}(a) represents a typical
implementation of a holographic immunoassay \cite{zagzag2020holographic,snyder2020holographic}
with antibodies
(IgG) binding to protein A on the surface of a probe.
THC measurements on thousands of beads are
pooled into estimates for the
population-averaged diameter, $d_0$,
typically with nanometer precision.
This value is compared with the mean probe-bead
diameter after
incubation, $d_p$, to estimate the change in diameter,
$\Delta d_p = d_p - d_0$.
The observed diameter change is related to the
thickness of the molecular-scale coating
\cite{snyder2020holographic,altman2020interpreting}
and therefore
to the concentration of the target analyte
\cite{snyder2020holographic}.
THC does not rely on fluorescent labeling
and thus eliminates
the materials and processing required for
fluorescence-based readout techniques.
This in turn reduces the cost, complexity and
time required to develop and deploy
bead-based molecular binding assays.

Although THC measurements yield results
with reproducibly high precision \cite{moyses2013robustness},
run-to-run variations can introduce systematic
offsets in estimated parameters
\cite{snyder2023aberration}
that reduce accuracy in $\Delta d_p$ and therefore
degrade the sensitivity and limit-of-detection
of the assay.
Here, we introduce a class of colloidal reference beads
whose properties are optimized
to detect and mitigate systematic errors
in THC measurements and thus
to optimize the repeatability and reproducibility of
all classes of measurements based on THC,
including molecular binding assays.
We introduce all-optical techniques to characterize
the dense polymer brush that stabilizes the reference
beads, including a method to measure the specific volume
of polymers and a THC-based method to measure the
grafting density of those polymers on the reference
beads' surfaces.
We demonstrate the utility of reference beads
for holographic assays by performing a
multiplexed holographic immunoassay for IgG using inert
reference beads as an internal negative control.
The success of this assay serves to validate
the effective-medium interpretation of holographic
characterization data for inhomogeneous
colloidal materials.
In addition to advancing holographic binding assays
as an analytical platform, the techniques introduced
for this study will be useful in any context involving
quantitative characterization of colloids.

\begin{figure}
    \centering
    \includegraphics[width=0.9\columnwidth]{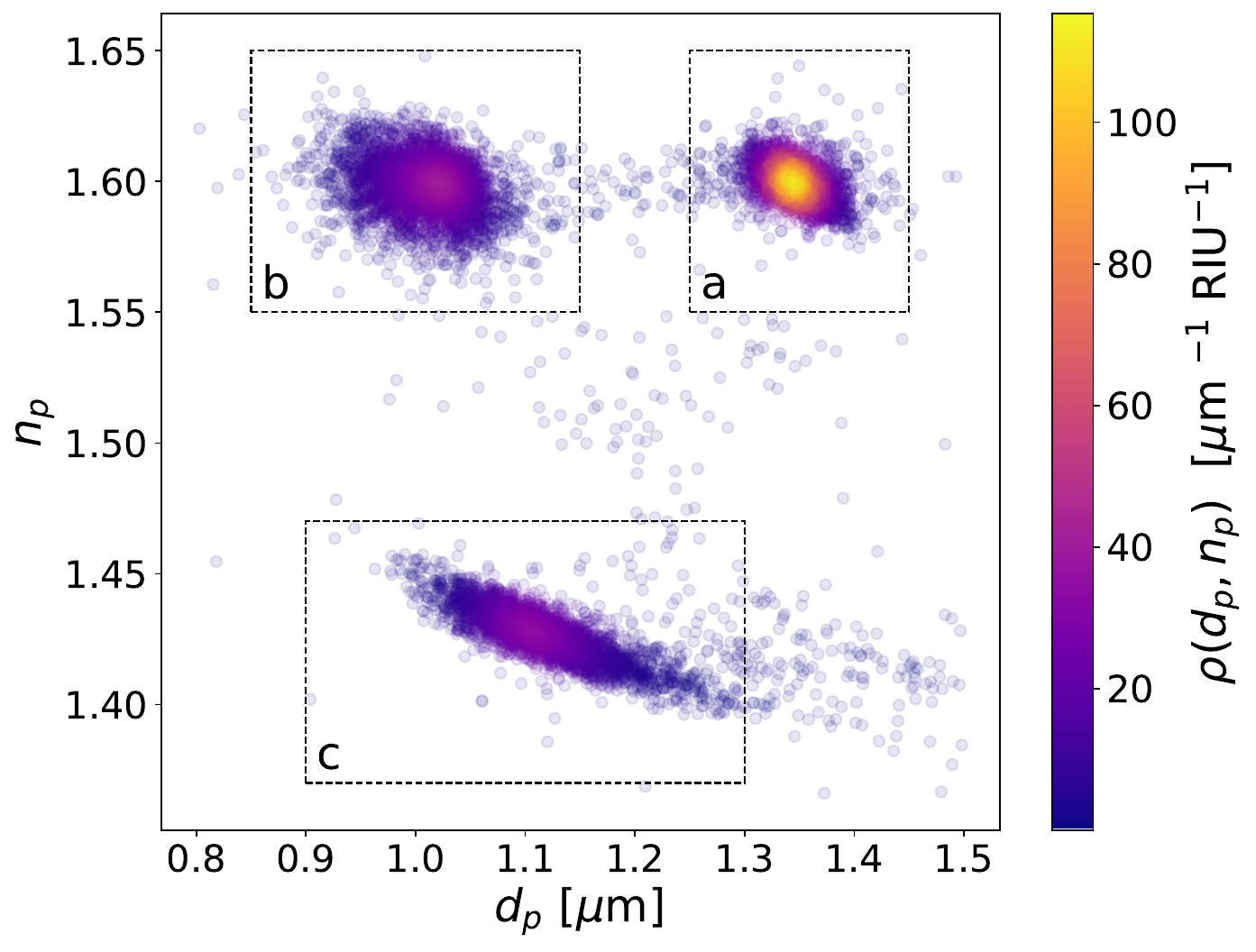}
    \caption{Holographic particle characterization data for
    a mixture of (a) reference beads, (b) immunoassay probe beads
    with PS substrates and (c) immunoassay probe beads with
    silica substrates, all in the same dispersion at an overall
    concentration of \qty{3e6}{beads\per\milli\liter}. Each of the
    \num{11616} points represents the diameter, $d_p$, and refractive index, $n_p$, of a single particle.
    Points are colored by the density of observations, $\rho(d_p, n_p)$.
    Each bead type occupies a distinct region of the
    $d_p$- $n_p$ plane allowing for multiplexed assays.}
    \label{fig:distribution}
\end{figure}

\section{Materials and Methods}

\subsection{Multiplexed assay kit with negative controls}
\label{sec:testkit}

The test kit used to perform multiplexed holographic molecular
binding assays for immunoglobulins
is a colloidal dispersion composed of
three types of colloidal beads:
\qty{1}{\um}-diameter polystyrene (PS) probe beads precoated with protein A (Bangs Laboratories, catalog no.~CP02000, lot no.~14540) \cite{snyder2020holographic};
\qty{1}{\um}-diameter silica probe beads
(General Engineering \& Research)
custom coated with protein A;
and \qty{1.3}{\um}-diameter PS reference beads
(Thermo Fisher Scientific, catalog no.~5130A, lot no.~172008)
custom coated with poly(ethylene oxide) (PEO).
Protein A enables functionalized beads to bind antibodies to their surfaces
\cite{snyder2020holographic}.
The PEO-coated PS beads are designed to inhibit
macromolecular binding
and therefore serve as negative controls
for binding-induced changes in the functionalized test beads.

The PS test beads are coated with protein A binding sites by the
manufacturer.
The silica probe beads are functionalized with protein A using the a PolyLink Protein Coupling Kit (Bangs Laboratories, catalog no.~PL01N).
The PEO-coated PS references beads are prepared according
to the protocol in Sec.~\ref{sec:referencebeadfunctionalization}.
These latter beads not only lack specific binding sites,
but are passivated by a dense polymer brush that
sterically stabilizes their surfaces.

Probe beads and reference beads are dispersed at 1:1:1 stoichiometry
in antibody binding buffer consisting of
\qty{50}{\milli\Molar} sodium borate buffer
prepared with boric acid
(\qty{99.5}{\percent}, Sigma-Aldrich, catalog no.~B0394, lot no.~SLBM4465V)
and \ce{NaOH}
(\qty{98}{\percent}, Sigma-Aldrich, catalog no.~S8045, lot no.~091M01421V)
in deionized water
(\qty{18.2}{\mega\ohm\cm}, Barnstead Millipure).
The buffer is adjusted to pH~\num{8.2} with the addition of dilute \ce{HCl} (\qty{38}{\percent}, Sigma-Aldrich, catalog no.~H1758) \cite{fishman2019antibody}.
Bovine serum albumin (BSA) (Sigma Aldrich, catalog no.~A4503) is added at \qty{0.01}{\percent~w/v} to
inhibit nonspecific binding by blocking any bare regions on the surfaces of the probe beads.
The overall concentration of beads is adjusted to \qty{3e6}{beads\per\milli\liter}
for compatibility with THC.

\subsection{Total Holographic Characterization of probe beads and reference beads}

THC measurements are performed with a commercial instrument
(xSight, Spheryx).
Each measurement involves transferring \qty{30}{\micro\liter}
of the fluid sample into one of the reservoirs in a
dedicated microfluidic chip (xCell8, Spheryx).
The measurement proceeds automatically,
with a selected portion
of the fluid in the reservoir being transported
through the instrument's observation
volume in a pressure-driven flow.
THC analysis provides
a record of the diameter, refractive index, and morphology
of each particle in the measured sample, together
with uncertainties in those values.
THC can be applied to beads ranging in
diameter from \qty{500}{\nm} to \qty{10}{\um}
dispersed in fluid media at concentrations
ranging from \qty{e3}{beads\per\milli\liter}
to \qty{e7}{beads\per\milli\liter}.
Figure~\ref{fig:distribution} shows the
results from one such measurement, with each plot symbol representing the diameter, $d_p$, and refractive index, $n_p$, one particle
in a stoichiometric mixture of probe beads and reference beads.
A \qty{1}{\micro\liter} volume of this sample
yields results for roughly \num{1000} beads
of each type, indicating an overall concentration of \qty{3e6}{beads\per\milli\liter}.
The data points in Fig.~\ref{fig:distribution}
are colored by the density of observations,
$\rho(d_p, n_p)$ and form three clusters,
one for each population of particles.
Regions of interest indicated in
Fig.~\ref{fig:distribution} identify
each of these three populations and are
used to assess variations in the
properties of those populations from
run to run.

THC measurements on bare PS
substrate beads yields population-averaged values for the diameter,
$d_0 = \qty{1.3408(2)}{\um}$, and the refractive index,
$n_0 = \num{1.6019(1)}$.
Consistent values are obtained
when the beads are dispersed
in DI water
or in \qty{1}{\mM} \ce{NaCl}.
The uncertainty in the last digit is the
standard error of the mean for
the entire set of measurements, and
reflects the precision with which changes in
particle diameter and refractive index can be resolved.

The standard error in the mean diameter is much smaller than
the standard deviation,
$\sigma_d = \qty{0.028}{\um}$,
which reflects the \qty{2}{\percent} polydispersity
in the beads' underlying size distribution.
Pooling independent measurements accounts for run-to-run
variations that are specified by the
instrument's manufacturer to be as large as \qty{5}{\nm} in the diameter
and \num{0.003} in the refractive index.

\subsection{Reference bead functionalization}
\label{sec:referencebeadfunctionalization}

As illustrated schematically in Fig.~\ref{fig:beads}(b),
the reference beads introduced in this study for THC measurements
consist of monodisperse polystyrene (PS) spheres
whose surfaces
are sterically stabilized against physisorption of macromolecules
by a dense coating of poly(ethylene oxide) (PEO).
Polystyrene beads are suitable substrates
both because of their commercial availability
and also because their
high refractive index minimizes the influence of
any unintended adsorbates on the their light-scattering properties
\cite{altman2020interpreting}, which is a desirable feature
for reference particles.

The protocol used to coat PS beads with a dense brush of PEO is  described in Ref.~[\onlinecite{oh2015peo}].
Commercial PS beads  are swollen by adding tetrahydrofuran (THF) to their aqueous dispersion and are
incubated with a PS-b-PEO diblock copolymer.
The hydrophobic PS blocks
dissolve in the swollen PS spheres and
are anchored in place by evaporating the THF
to deswell the beads.
This protocol leaves the hydrophilic PEO blocks
exposed on the beads' surfaces.

Reference beads are
prepared by dispersing
\qty{1.4}{\um}-diameter PS beads
(Thermo Fisher Scientific, catalog no.~5130A, lot no.~172008)
at \qty{10}{\percent~w/v} in
a solution composed of
\qty{140}{\micro\liter} poly(styrene-b-ethylene oxide) (PS-b-PEO) solution (catalog no.~P1807A-SEO, Polymer Source, Inc.),
\qty{90}{\micro\liter} DI water,
and \qty{160}{\micro\liter} THF
at room temperature.
The block copolymer consists of
\qty{3.8}{\kilo\dalton} PS covalently linked to \qty{34}{\kilo\dalton} PEO.
This mixture is placed on a horizontal shaker at \qty{900}{rpm} for \qty{1.5}{\hour}.
After incubation, THF is removed from the solution by evaporation at room temperature over the course of \qty{2}{\hour}.
Excess polymer is removed by
washing the particles three times in deionized
water.
Each washing cycle involves
centrifuging the dispersion at \qty{6500}{rpm} for \qty{5}{\minute} to concentrate the beads, removing the supernatant,
and redispersing the beads in deionized water.

The tethered PEO block has a contour length of
roughly $\ell = \qty{216}{\nm}$,
assuming an incremental length of
\qty{0.28}{\nm\per monomer}
\cite{lee2008molecular, oesterhelt1999single}
and an expected
radius of gyration
of $R_g = \qty{9.1}{\nm}$ \cite{zheng2023dna}.
The stabilizing properties of the
molecular brush formed by these
molecules depend on the
molecules' specific
volume
and their grafting density on the
surface,
both of which can be measured optically.

\subsection{Optical specific volume of PEO}
\label{sec:specificvolume}

\begin{figure*}
    \centering
    \includegraphics[width=0.8\textwidth]{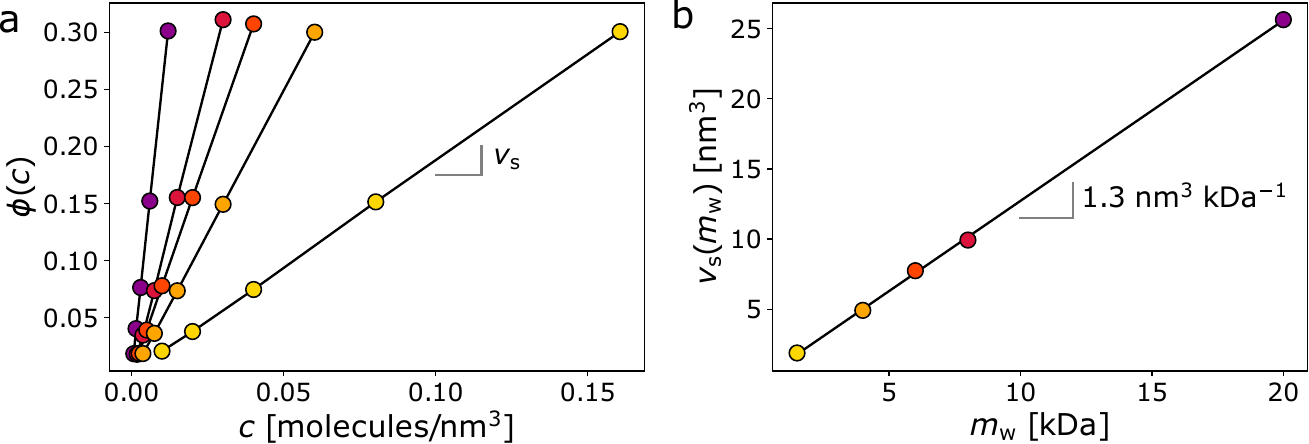}
    \caption{(a) The volume fraction, $\phi(c)$, occupied by PEG molecules in solution as a function of concentration, $c$, for five molecular weights,
    estimated from refractive index measurements using
    Eq.~\eqref{eq:specificvolume}. The slope of each line gives the optical specific volume, $v_s$ for each molecular weight, $m_w$. (b) The optical specific volume of each PEG molecule as a function of molecular weight obtained from (a). The slope,
    \qty{1.3}{\cubic\nm\per\kilo\dalton} can be used
    to estimate $v_s$ for other molecular weights.}
    \label{fig:specificvolume}
\end{figure*}

The optical specific volume of a polymer
is the volume, $v_s$,
that a single molecule
occupies in solution
as gauged by the molecule's influence
on the solution's optical properties
\cite{heller1966application}.
This value does not include solvation
effects and is largely independent of
the molecule's conformation.
The optical specific volume
is especially useful for the insights
it offers into
the relative contribution of each
molecular species to the refractive
index of a heterogeneous medium.
In that respect, it is complementary
to standard $dn/dc$ analysis \cite{striegel2017specific},
in which the refractive index increment is used
to measure molecular weight rather than
molecular volume.

The measured refractive index of a polymer solution, $n$,
is related to the volume fraction, $\phi$,
of polymer molecules in solution
through Maxwell Garnett effective-medium
theory
\cite{markel2016introduction}:
\begin{subequations}
\label{eq:specificvolume}
\begin{equation}
    L\left(\frac{n}{n_m}\right)
    =
    \phi \, L\left(\frac{n_1}{n_m}\right) ,
\end{equation}
where $n_1$ is the intrinsic refractive index of the polymer, $n_m$ is the refractive index of the fluid medium and
\begin{equation}
    L(m) = \frac{m^2 - 1}{m^2 + 2}
\end{equation}
is the Lorentz-Lorenz factor.
The volume fraction, in turn, is proportional to the
polymer's concentration, $c$,
\begin{equation}
    \phi = c \, v_s,
\end{equation}
\end{subequations}
where $v_s$ is the specific volume of a single polymer
molecule.
Equation~\eqref{eq:specificvolume} therefore describes a protocol
for measuring the optical specific volume of a polymer
based on measurements of $n(c)$.

The data in Fig.~\ref{fig:specificvolume} were obtained for
solutions of poly(ethylene glycol) (PEG) of various molecular
weights in \qty{5}{\mM} sodium phosphate buffer at pH~7.
The refractive index of each solution is measured
with an Abbe refractometer (Edmund Optics, model~52-975)
at a vacuum wavelength of \qty{589}{\nm}
and is used to estimate
the volume fraction of polymer in
solution according to
Eq.~\eqref{eq:specificvolume}
using
$n_m = \num{1.3325(5)}$ and
$n_1 = \num{1.466(4)}$
\cite{ingham1965refractive,ottani2002densities, weissler1947sound}.
The five data sets in Fig.~\ref{fig:specificvolume}(a)
show results for five different mean molecular weights:
\qty{1.5}{\kilo\dalton} (Fluka Analytical, catalog no.~81210),
\qty{4}{\kilo\dalton} (bioWorld, catalog no.~714224),
\qty{6}{\kilo\dalton} (Millipore, catalog no.~528877),
\qty{8}{\kilo\dalton} (bioWorld, catalog no.~705631)
and \qty{20}{\kilo\dalton} (Fluka Analytical, catalog no.~95172).
Each sample is prepared at \qty{40}{\percent~w/v}
and is sequentially diluted by factors of 2
to obtain $\phi(c)$.
The slope of each linear trend yields $v_s$ for PEG
at the associated molecular weight.
We assume that the same value for the
optical specific volume can be used to
interpret optical measurements at other
wavelengths,
including THC measurements.

The refractometry data plotted in Fig.~\ref{fig:specificvolume}(b)
confirm that the specific volume of PEO, $v_s$,
scales linearly with molecular weight, $m_w$,
with a slope $d v_s/d m_w = \qty{1.308(4)}{\cubic\nm\per\kilo\dalton}$.
This fundamental property of PEO
appears not to have been reported
previously and is useful for estimating the
specific volume of PEG and PEO at arbitrary molecular weights.
Applying it to the PS-b-PEO diblock
copolymer used
to functionalize the reference beads
yields a value for the specific volume of
the \qty{34}{\kilo\dalton} PEO block:
$v_s = \qty{44.5(1)}{\cubic\nm}$.

\subsection{THC measurement of PEO grafting density}
\label{sec:alloptical}

As depicted schematically in Fig.~\ref{fig:beads}(b),
our reference beads are coated with a dense brush
of PEO.
The quality of their functionalization is gauged
by the grafting density, $\Gamma_c$, of PEO molecules on the surface
of the PS bead.
Here, we introduce a method to measure the grafting density
based on THC measurements of the beads
before and after functionalization.

Differential THC measurements yield the apparent
increase, $\Delta d_p$, in the the beads'
average diameter after functionalization.
The observed diameter shift does not necessarily
correspond to twice the coating thickness, $a_c$,
because the refractive index of the coating,
$n_c$, generally does not match that
of the substrate bead, $n_0$.
Instead, $\Delta d_p$ is related to $a_c$
by \cite{altman2020interpreting}
\begin{subequations}
\label{eq:graftingdensity}
\begin{equation}
    \label{eq:altmangrier}
    \Delta d_p
    =
    2 a_c \, \frac{n_c - n_m}{n_0 - n_m}.
\end{equation}
The refractive index of the coating depends
on the volume fraction, $\phi_c$, of macromolecules
in the coating through the Lorentz-Lorenz relation,
\begin{equation}
    L(m_c)
    =
    \phi_c \, L(m_1) ,
\end{equation}
where $m_c = n_c / n_m$ is the relative refractive
index of the coating and
$m_1 = n_1 / n_m$ is the relative refractive index
of a single macromolecule.
The volume fraction, in turn, depends on the
grafting density,
\begin{equation}
    \phi_c
    =
    \frac{v_s \Gamma_c}{a_c},
\end{equation}
\end{subequations}
where $v_s$ is the optical specific volume
of the grafted PEO molecule discussed in
Sec.~\ref{sec:specificvolume}.

\begin{figure}
    \centering
    \includegraphics[width=0.9\columnwidth]{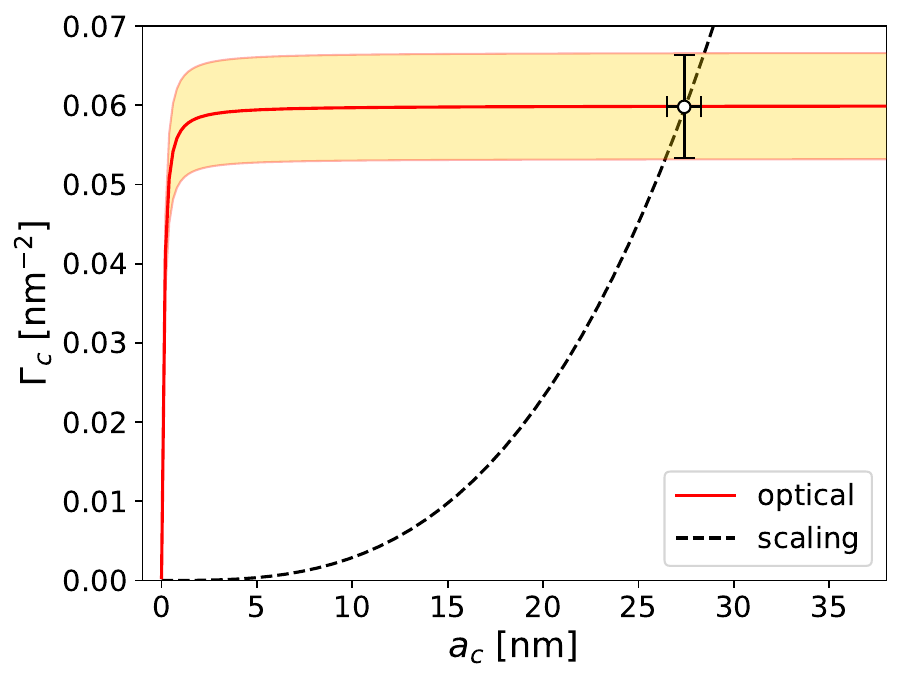}
    \caption{Possible values of the grafting density, $\Gamma_c$, and the coating thickness, $a_c$, of the PEO brunch on PS substrate beads consistent with THC measurements of
    the diameter increase, $\Delta d_p$,
    due to grafting. The dashed
    curve is the scaling prediction
    for a grafted polymer brush.}
    \label{fig:grafting}
\end{figure}

Equation~\eqref{eq:graftingdensity} uses
a THC measurement of $\Delta d_p$ to impose
three constraints on four characteristics
of the coating: $\Gamma_c$, $a_c$, $n_c$ and $\phi_c$.
This set of relationships
can be expressed as an overall constraint on the grafting density, $\Gamma_c$, and
layer thickness, $a_c$, conditioned on
the measurement of $\Delta d_p$:
\begin{equation}
    \label{eq:graftingdensityconstraint}
    \Gamma_c(a_c | \Delta d_p)
    =
    \frac{x}{v_s L(m_1)}
    \frac{a_c^2 + x a_c}{\frac{3}{4} a_c^2 + x a_c + x^2}
\end{equation}
where $x = \Delta d_p  (m_0 - 1) / 4$.

Equation~\eqref{eq:graftingdensityconstraint} applies
without modification when macromolecules are grafted
or physisorbed directly to the surface of the substrate
bead.
The PS block of the copolymer brush, however,
is embedded in the substrate bead.
This increases
the bead's diameter by an amount that depends on
the grafting density:
\begin{equation}
    \Delta d_0(\Gamma_c)
    \approx
    3 \frac{m_\text{PS}}{\rho_\text{PS}} \, \Gamma_c ,
\end{equation}
where $m_\text{PS} = \qty{3.8}{\kilo\dalton}$
is the mass of the PS block and
$\rho_\text{PS} = \qty{1.05e3}{\kg\per\cubic\meter}$
is the mass density of polystyrene.
Equation~\eqref{eq:graftingdensityconstraint}
then must be solved for $\Gamma_c$
self-consistently using
\begin{equation}
    x = \frac{m_0 - 1}{4}
    \left[ \Delta d_p - \Delta d_0(\Gamma_c)
    \right] .
\end{equation}

Figure~\ref{fig:grafting} presents
$\Gamma_c(a_c | \Delta d_p)$ for the
\qty{34}{\kilo\dalton} PEO brush used to
functionalize our reference beads,
conditioned on THC measurements of
$\Delta d_p$.
THC measurements are performed in
\qty{1}{\mM} \ce{NaCl} solution
for consistency with electrokinetic
characterization measurements
discussed in Sec.~\ref{sec:electrokinetic}.
Each measurement is performed seven times to minimize the influence of
run-to-run instrumental variations and
the results are pooled.
This procedure
yields $\Delta d_p = \qty{3.3(3)}{\nm}$.
The other material parameters in
Eq.~\eqref{eq:graftingdensityconstraint} are
the refractive indexes
$n_1 = \num{1.470(5)}$ and
$n_m = \num{1.340(1)}$, which
are appropriate for PEO and water, respectively,
at the wavelength of light
used for THC,
and $v_s = \qty{44.5(1)}{\cubic\nm}$, which is obtained in
Sec.~\ref{sec:specificvolume}.
The shaded region in Fig.~\ref{fig:grafting}
reflects the uncertainties in these values.
These results self-consistently
account for an increase in the substrate bead diameter
up to
$\Delta d_0 = \qty{0.7(1)}{\nm}$ due to incorporation
of the PS blocks.

The dashed curve in Fig.~\ref{fig:grafting}
is the predicted scaling relationship
\cite{doi2013soft},
\begin{equation}
\label{eq:scaling}
    a_c(\Gamma_c)
    =
    \left(
    \frac{1}{6} \ell^2 v_s \, \Gamma_c
    \right)^{1/3} ,
\end{equation}
between the thickness of a
Gaussian polymer
brush and its grafting density.
Requiring consistency between this prediction
and the optical constraint
from Eq.~\eqref{eq:graftingdensityconstraint}
yields
$\Gamma_c = \qty{0.060(7)}{\per\square\nm}$
for the grafting density and
$a_c = \qty{27(1)}{\nm}$
for the effective coating thickness.
Comparable results for $\Gamma_c$ have been reported
for the same grafting
procedure on similar beads using orthogonal
measurement techniques
\cite{oh2015peo,zheng2023dna}.

The estimated value of $\Gamma_c$ is consistent
with a mean separation between tethered
PEO molecules of
\qty{4}{\nm}, which is
is smaller than the individual molecules'
radius of gyration, $R_g = \qty{9.1}{\nm}$.
The coating therefore is sufficiently dense
for the grafted polymers to form a brush.
The separation
also is smaller than the dimensions of
typical
target analytes for molecular binding assays,
which contributes to the stability of the PEO-coated
reference beads against physisorption.

\subsection{Electrokinetic characterization of reference beads}
\label{sec:electrokinetic}

Electrokinetic characterization measurements
are performed on the PS reference beads
using a Zetasizer Nano
(Malvern Panalytical)
both
before and after functionalization with PEO.
Unlike THC measurements, which can be
performed in any fluid medium,
electrokinetic characterization
requires the particles
to be suspended in \qty{1}{\mM} \ce{NaCl}
solution.
The bare beads
yield a zeta potential, $\zeta_p = \qty{-81}{\milli\volt}$,
and electrophoretic mobility, $\mu_e = \qty{-6.3}{\um\cm\per\volt\per\second}$,
that are consistent with expectations
\cite{hidalgo1996electrokinetic}
for polystyrene sulfonate beads.
The corresponding values after coating with PEO,
$\zeta_p = \qty{-2.1}{\milli\volt}$
and $\mu_e = \qty{-0.16}{\um\cm\per\volt\per\second}$,
show a forty-fold reduction in effective
surface charge, which correspondingly reduces
the beads' interactions with charged species
in solution.
Fitting these data to the standard model
for electrophoretic mobility of ``fuzzy'' colloids
\cite{hill2005exact,zheng2023dna}
yields a brush thickness of
$a_c = \qty{34(2)}{\nm}$
\cite{zheng2023dna}.
This value is larger than the all-optical
result reported in Sec.~\ref{sec:alloptical},
which may differences in the assumptions
underlying the models used to analyze the
two types of measurements.
Most notably, Eq.~\eqref{eq:graftingdensity}
does not account for the radial density gradient
in the polymer brush and therefore tends
to underestimate $a_c$.

\begin{figure*}
    \centering
    \includegraphics[width=0.8\textwidth]{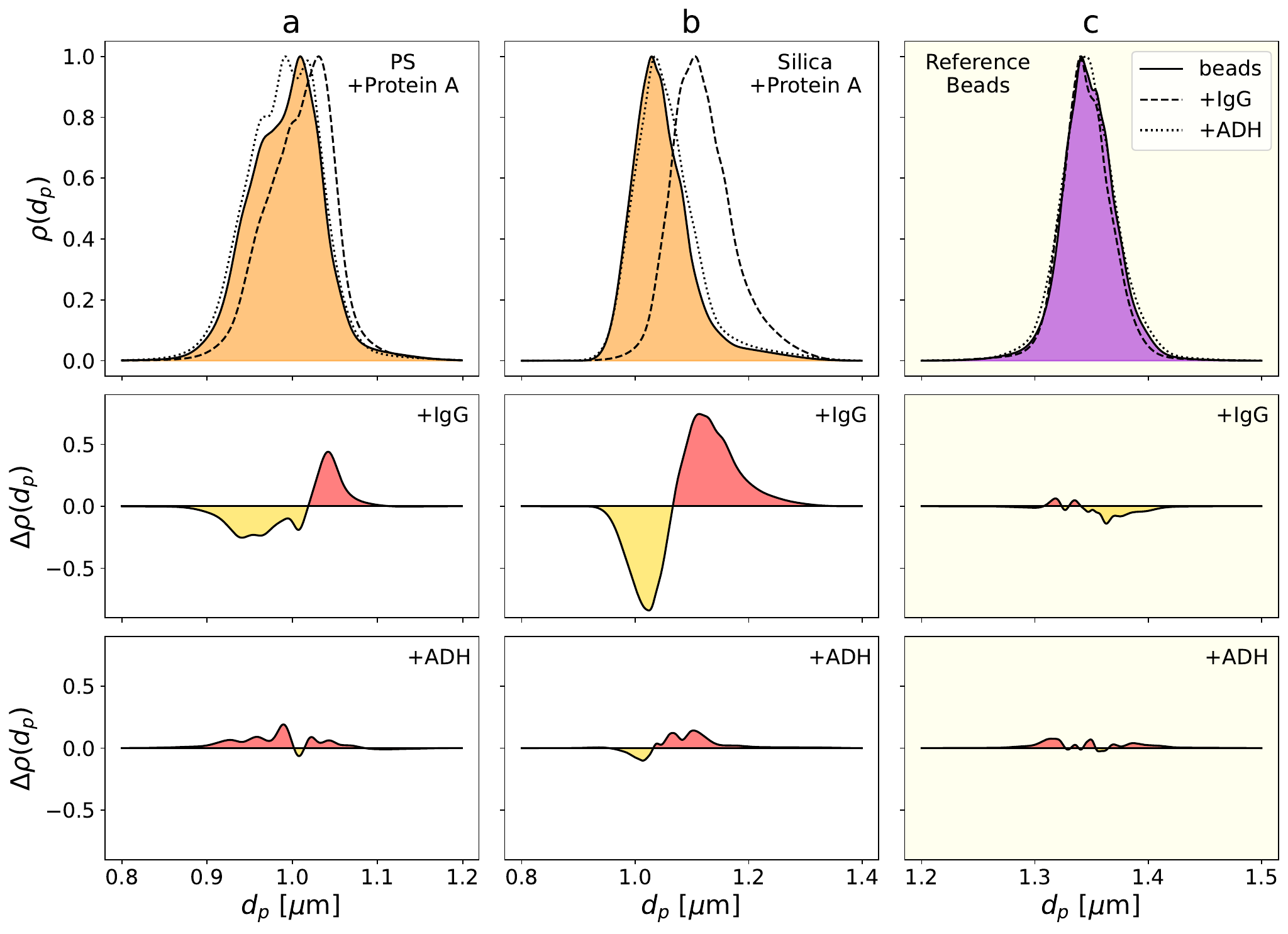}
    \caption{(top row) THC measurements of the diameter distribution, $\rho(d_p)$,
    for (a) PS probe beads, (b) silica probe beads and (c) PS reference beads before
    and after incubation with IgG
    or ADH.
    (middle row)
    Differences, $\Delta \rho(d_p)$, between the
    distributions before and after incubation distributions with \qty{50}{\ug\per\milli\liter} IgG showing clear positive responses from
    the probe beads and no response
    from the reference beads.
    (bottom row) Control measurements of
    $\Delta \rho(d_p)$
    with \qty{50}{\ug\per\milli\liter}
    ADH show no response from
    either probe or reference beads.}
    \label{fig:orig_kde}
\end{figure*}

\begin{table}
\sisetup{separate-uncertainty=false}
\caption{Mean bead diameter, $d_0$, and observed increase in bead diameter,
$\Delta d_p$, after incubation with \qty{50}{\micro\gram\per\milli\liter} of
either IgG or ADH. Probe beads functionalized with protein A respond to
IgG whereas PEO-coated reference beads do not. Negative
control measurements with ADH elicit no response from any of the beads.}
\begin{ruledtabular}
\begin{tabular}{lSS[table-format=2.0(1)]S[table-format=1.0(1)]}
     Bead Type & {$d_0$ [\si{\um}]} & \multicolumn{2}{c}{{$\Delta d_p$ [\si{\nm}]}} \\
     & & {IgG} & {ADH} \\
     \hline
     PS Probe & 0.9920(6) & 16(1) & -2(1) \\
     Silica Probe & 1.0436(8) & 71(2) & 8(2) \\
     PS Reference & 1.3476(4) & -2(1) & 0(1) \\ % -2.3(6) & -0.2(8) \\
\end{tabular}
\end{ruledtabular}
\label{tab:diameters}
\end{table}

\subsection{Holographic analyte binding measurements}
\label{sec:assayprocedure}

Label-free holographic binding assays are
performed at an analyte concentration of
\qty{2.5}{\milli\gram\per\milli\liter}
to mimic testing conditions at the lower end
of the physiologically relevant range of concentrations \cite{morell1976serum,snyder2020holographic}.
The same concentration of ADH is used as a
negative control to both to demonstrate that
the passivated probe beads are free from
nonspecific binding and also to demonstrate
that the reference beads are inert.

The assay is initiated by adding a
\qty{4}{\micro\liter} aliquot of the analyte solution
to \qty{196}{\micro\liter}
of the multiplexed assay kit dispersion.
Adding the analyte to the test solution effectively
dilutes the analyte to a concentration of
\SI{50}{\micro\gram\per\milli\liter}.
The system is allowed to incubate at
room temperature for \qty{45}{\minute}
under gentle agitation on a rotary
platform (MS 3 basic, IKA).
Three \qty{30}{\micro\liter} samples
are then transferred into
separate reservoirs on microfluidic
chips for THC analysis in triplicate.
Each such measurement yields a distribution
of holographically detected particle
properties such as the example in
Fig.~\ref{fig:distribution}.

The three types of beads in the mixed dispersion
give rise to distinctive clusters of points in the
$d_p$-$n_p$ plane, which allows for straightforward
data segmentation.
The dashed boxes in Fig.~\ref{fig:distribution}
reflect the sampling regions used for this study.
The data clusters associated with each
bead type are then compared from run to run with
Kolmogorov-Smirnov tests to ensure consistency.
If the replicated measurements are found to be
consistent, the results for each
bead type are pooled.
The mean diameter, $d_p$, then is calculated for
each bead type, along with the standard
error in the mean.
These values then are compared with
values obtained for the same beads in the
same buffer before incubation to obtain
the differences, $\Delta d_p$, for each
bead type.
These differences then can be used to assess whether
the target analyte was present in the sample
and to infer its concentration \cite{zagzag2020holographic,snyder2020holographic}.

Typical data for a label-free bead-based holographic
molecular binding assay are plotted in Fig.~\ref{fig:orig_kde}
and are summarized in Table~\ref{tab:diameters}.
The top row in Fig.~\ref{fig:orig_kde} shows pooled
probability densities for the particle diameters, $\rho(d_p)$,
for each of the three bead types.
Distributions are normalized to unity peak height to facilitate
comparison of the shapes of the distributions.
Results are shown for the stock beads before incubation
(shaded) and for the beads after incubation with
IgG and with ADH. The middle and bottom rows show changes
in the distributions after incubation.
The PS probe beads show a clear response to incubation with
IgG and very little response to incubation with ADH.
Silica probe beads respond more strongly to IgG and
just as weakly to negative control measurements with ADH.
The reference beads developed for this study show no response
either to IgG or to ADH.

\section{Results}
\label{sec:results}

\subsection{Optical specific volume of PEO}

The measurement technique described in Sec.~\ref{sec:specificvolume} yields
\begin{equation}
    \frac{d v_s}{d m_w}
    =
    \qty{1.308(4)}{\cubic\nm\per\kilo\dalton}
\end{equation}
for the differential specific volume of PEO and PEG.
When multiplied by the molecular weight, this value represents the intrinsic volume of the macromolecule
associated with its interaction with light,
independent of conformation or solvent interactions.

\subsection{All-optical measurement of macromolecular grafting density}

Holographic measurements of the diameter
of PEO-coated beads described in
Sec.~\ref{sec:alloptical}
are combined with
refractometry measurements of the macromolecular specific volume to obtain a constraint condition
between the grafting density of macromolecules
and the thickness of the macromolecular coating.
Requiring consistency with theoretical predictions
from Eq.~\eqref{eq:scaling} yields
the grafting density of PEO on the reference beads prepared for this study:
\begin{equation}
    \Gamma_c = \qty{0.060(7)}{\per\square\nm} .
\end{equation}

\subsection{Multiplexed molecular binding assay with
integrated negative control}

\begin{figure*}
    \centering
    \includegraphics[width=0.9\textwidth]{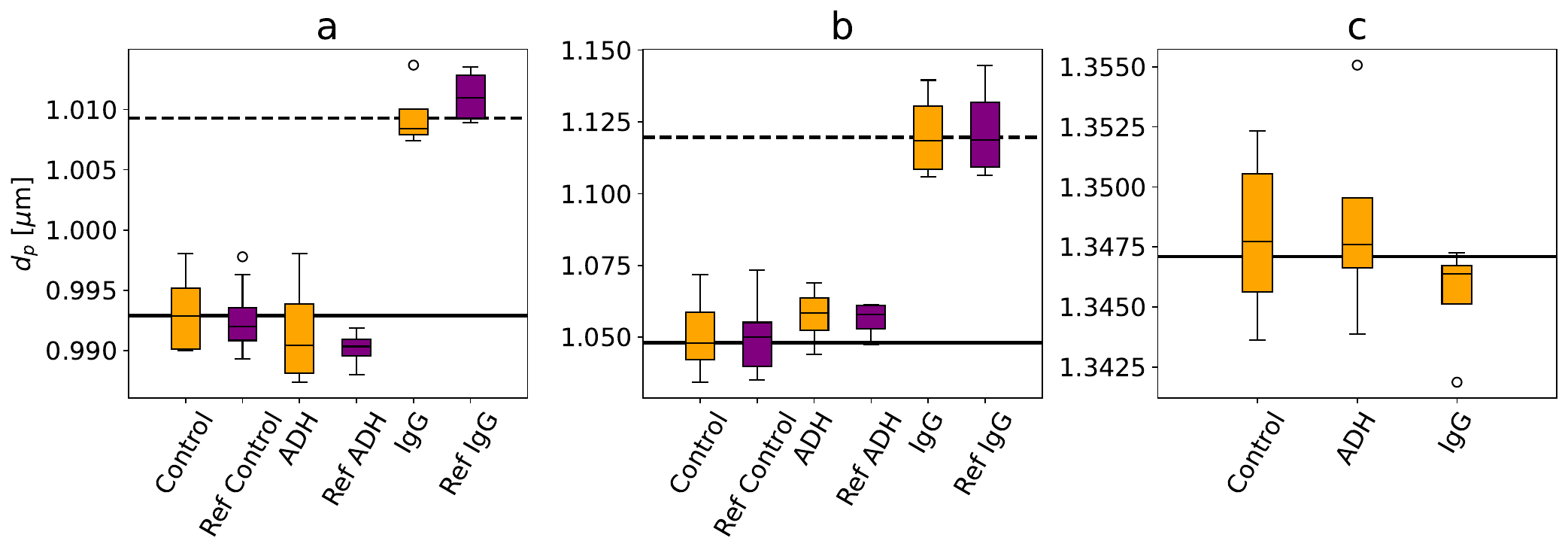}
    \caption{Box and whisker plots showing bead diameter distributions in multiplexed holographic immunoassays before (orange) and after (purple) correcting for run-to-run variations with reference-bead standards.
    (a) PS test beads functionalized with protein A. (b) Silica test beads functionalized with protein A. (c) PEO-passivated PS reference beads. Each result is
    pooled from at least three replicated measurements and includes holographic characterization data from at least \num{5000} beads. Control measurements are
    performed on the assay kit before incubation with
    analyte solutions.}
    \label{fig:immunoassay}
\end{figure*}

Figure~\ref{fig:immunoassay} reports the
results of a multiplexed immunoassay performed with the
test kit described in Sec.~\ref{sec:testkit}.
This test kit performs two independent assays
for antibodies such as IgG, and provides
internal negative controls using the reference
beads developed in Sec.~\ref{sec:referencebeadfunctionalization}
and characterized in
Sec.~\ref{sec:alloptical} and
Sec.~\ref{sec:electrokinetic}.
The three classes of beads are codispersed in the
same test kit and therefore are brought into
equilibrium with the same analyte solution
following the procedure in Sec.~\ref{sec:assayprocedure}.
The mean diameter, $d_p$, of each population of beads
is determined
using Total Holographic Characterization
on statistical samples of \num{1000} beads of each
type both before
and after incubation.
These measurements are repeated in triplicate
and the results are pooled.
The (orange) box-and-whisker plots
in Fig.~\ref{fig:immunoassay} report the
population-averaged values of $d_p$ for
control measurements before incubation,
negative control measurements
with ADH, and test measurements with IgG
for each of the three classes of beads.
These assays report a positive response
with IgG in both the PS test beads
($\Delta d_p = \qty{16(1)}{\nm}$)
and the silica test beads
($\Delta d_p = \qty{71(2)}{\nm}$).

The observed sensitivity of the PS-bead assay
agrees with previously reported results
\cite{snyder2020holographic}
for this type of test bead at this
analyte concentration, although
the improved measurement protocol provides
a four-fold reduction in uncertainty.
The silica test beads have a four-fold larger
response in the same analyte.
This difference can be attributed to the lower refractive index of the silica
substrate beads and provides an experimental
confirmation of the predicted dependence of an assay's response on the substrate bead's composition \cite{altman2020interpreting}.

As reported in Fig.~\ref{fig:immunoassay}(c),
the inert PEO-coated PS beads show no statistically significant response to either ADH or IgG, confirming
their utility as built-in negative controls for holographic binding assays.
From run to run, however, the measured values of
the mean bead diameter can vary by as much
as \qty{5}{\nm}, possibly due to systematic
effects such as manufacturing variations in
the microfluidic cells' optical properties.
Such systematic effects might also influence
the results for the probe beads.
To assess and potentially mitigate such
effects, we compute the deviation, $\delta d_p$,
of the measured diameter of the reference beads
from the nominal ground-truth value reported in
Table~\ref{tab:diameters}
for each run.
We then subtract the same deviation from the
measured diameters of the other types of beads
in the same run.
This naive correction yields the
(purple) reference-corrected results in
Fig.~\ref{fig:immunoassay}.
For this set of measurements, reference-based
corrections do not significantly change the
mean bead diameters obtained by pooling replicated
measurements.
This suggests that errors are dominated by statistical
uncertainty rather than systematic instrumental errors.
The presence of the reference beads in this multiplexed
assay therefore helps to validate the assay by
minimizing the possibility that the detection of
IgG resulted from a false positive reading or that
the absence of a signal in the ADH assay resulted
from a false negative reading.

The two independent assays for IgG in the
multiplexed test kit should report consistent
results for the concentration of the target analyte
despite the four-fold difference in their
responses to analyte binding.
In both cases, the thickness, $a_c$,
of the molecular coating formed by bound IgG
should be comparable to the size of a single antibody.
The coating's effective refractive index, $n_c$,
depends on the areal density of binding sites
on the beads' surfaces and their fractional occupation
in equilibrium given the analyte's concentration
\cite{snyder2020holographic}.
If we assume that the two types of probe beads have
comparable areal densities of binding sites,
then Eq.~\eqref{eq:altmangrier} suggests that
$A \equiv \Delta d_p (n_0 - n_m) = 2 a_c (n_c - n_m)$
should have the same value in both assays.
The results,
\begin{subequations}
\begin{align}
    A_\text{PS} & = \qty{4.0(3)}{\nm} \\
    A_\text{silica} & = \qty{4.3(4)}{\nm},
\end{align}
\end{subequations}
indeed are consistent with each other
and support the conclusion from earlier reports \cite{zagzag2020holographic,snyder2020holographic,altman2020interpreting}
that effective-medium analysis of
holographic bead-based binding assays
accurately models macromolecular binding and therefore
yields accurate analyte concentrations
regardless of the composition
of the substrate beads.

\section{Conclusions and Discussion}

Total Holographic Characterization (THC) measures the diameters
of micrometer-scale colloidal beads with the precision and
accuracy needed to detect nanometer-scale macromolecules
binding to their surfaces.
The standard analysis for THC \cite{lee2007characterizing} treats
each bead as a homogeneous sphere.
THC data for inhomogeneous particles such as coated beads
can be interpreted with effective-medium theory \cite{markel2016introduction,odete2020role,snyder2020holographic,altman2020interpreting}
to obtain information about properties of the coating.
The present study uses this all-optical technique to achieve two goals:
(1) to measure the grafting density of a poly(ethylene oxide) (PEO) brush
on the surface of polystyrene (PS) beads, and
(2) to demonstrate a multiplexed label-free immunoassay
using the PEO-coated PS beads as internal negative controls.
The techniques developed for this proof-of-concept demonstration
can be applied in any context where macromolecular coatings form on colloids
and can be particularly useful for assessing properties of the coatings.

The two independent assays implemented by the multiplexed test
kit described in Sec.~\ref{sec:testkit} both use protein A to
bind IgG to the test beads' surfaces.
THC clearly distinguishes the two different types of test beads
from each other and from the reference beads both by size and
also by refractive index.
The two tests both respond strongly to the presence of IgG in the
analyte solution and show no significant
response to negative control measurements performed with ADH.
By performing parallel independent assays for the
same analyte, we are able to verify that the
quantitative assay results are consistent with each other
when interpreted with effective-medium theory.
This consistency provides additional validation
for the overall use of effective-medium theory to
interpret THC results of coated spheres.
More generally, the test kit used for a holographic
molecular binding assay can combine multiple classes of holographically distinguishable beads that are each functionalized for distinct target molecules.

Including inert reference beads in the test kits
provides an internal assessment of run-to-run variations
in differential and replicated THC measurements.
In reporting the grafting density of the stabilizing polymer
brush on these reference beads, the present study also reports
the differential specific volume of poly(ethylene oxide) (PEO),
which appears not to have been reported previously.
In addition to detecting systematic errors due to
instrumental run-to-run variations, THC analysis
of the reference beads can help to mitigate those
variations, thereby increasing the accuracy of
multiplexed assays.

Most broadly, the present study demonstrates the viability
of multiplexed assays based on holographic characterization
of multiple classes of test beads codispersed in the
same test kit.
The present implementation dilutes the analyte from
physiologically relevant concentrations down to levels
suitable for bead-based binding assays.
Other combinations of analyte volume and
initial bead concentration
can be used to provide greater sensitivity at the
expense of a more limited range of accessible concentrations.
Demonstrations of such concentration-optimized
multi-target multiplexed assays will be reported
separately.

\section*{Acknowledgments}

The authors acknowledge helpful conversations with
David Pine and Jatin Abacousnac.
This work was supported by the National Science Foundation
under award no.~DMR-2104837.
The Spheryx xSight used for this study was purchased
as shared instrumentation by the MRSEC program of
the NSF under award no.~DMR-1420073.

\section*{Disclosures}

DGG is a founder of Spheryx, Inc., which manufactures
xSight for Total Holographic Characterization, including
the instrument used for this study.

%\bibliography{referencebeads.bib}
%apsrev4-2.bst 2019-01-14 (MD) hand-edited version of apsrev4-1.bst
%Control: key (0)
%Control: author (8) initials jnrlst
%Control: editor formatted (1) identically to author
%Control: production of article title (0) allowed
%Control: page (0) single
%Control: year (1) truncated
%Control: production of eprint (0) enabled
%

\end{document}